\documentstyle[12pt,aaspp4,flushrt]{article}
\begin{document}

\title{Extreme X-ray Variability In The Narrow-Line QSO PHL~1092}

\author{Karl Forster and Jules P. Halpern}
\affil{Department of Astronomy, Columbia University,
538 West 120th Street, New York, NY 10027}
\affil{Electronic mail: karlfor@mikado.phys.columbia.edu}
\received{December 19th, 1995}
\accepted{March 1st, 1996}
\affil{To appear in The Astrophysical Journal}

\begin{abstract}

A {\it ROSAT\/} observation of the narrow-line \ion{Fe}{2} QSO
PHL~1092 shows rapid variability that requires an efficiency of at
least 0.13, exceeding the theoretical maximum for an accretion disk
around a non-rotating black hole.  Plausible explanations for this
high efficiency incorporate anisotropic emission and/or accretion onto
a rapidly rotating black hole, the latter recently suggested by Kwan
et al. as a mechanism for generating PHL~1092's strong \ion{Fe}{2}
lines by mechanical heating in an accretion disk.  The soft X-ray
luminosity of PHL~1092 had also increased by a factor of 21 over the
weak {\it Einstein} detection, to more than $5 \times
10^{46}$~ergs~s$^{-1}$.  Its photon spectral index of 4.2 is among the
steepest of any AGN.  These X-ray properties are characteristic of
narrow-line Seyfert~1 galaxies, of which PHL~1092 is evidently a very
luminous member.  Narrow-line QSOs also extend a significant
correlation between X-ray luminosity and X-ray spectral index which we
have found among a large sample of optically-selected, narrow-line
Seyfert~1 galaxies observed by {\it ROSAT}.
\end{abstract}


\section{Introduction}

Narrow-line Seyfert~1 galaxies (NLS1s, Osterbrock \& Pogge 1985;
Goodrich 1989) are defined by their optical emission-line ratios and
widths: [\ion{O}{3}]/H$\beta < 3$ and FWHM H$\beta~<~2000$~km~s$^{-1}$.  
NLS1s also tend to have strong permitted \ion{Fe}{2}, \ion{Ca}{2} and
\ion{O}{1}~$\lambda$8446 emission lines (Persson 1988), as well as
high-ionization lines that are typical of Seyfert~1 galaxies.  Their
high X-ray luminosities were first noted by Remillard et al. (1986)
and Halpern \& Oke (1987).  With {\it ROSAT}, NLS1s were discovered to
have unprecedented X-ray properties (but first see Remillard et
al. 1991).  Their X-ray spectra are much softer than those of ordinary
Seyfert~1 galaxies (Brandt et al. 1994; Boller, Brandt, \& Fink 1996;
Pounds, Done, \& Osborne 1995), and they display rapid,
large-amplitude variability as well as extreme long-term changes
(Boller et al. 1993; Brandt, Pounds, \& Fink 1995; Grupe et
al. 1995a,b).  New members of this class found in the {\it ROSAT\/}
All-Sky Survey by Moran, Halpern, \& Helfand (1996) all have
\ion{Fe}{2} emission to some degree. The latter paper further argues
the need for a NLS1 class.

It has long been known that there are high-luminosity analogs of this
class, the prototype of which is I~Zw~1 (e.g., Phillips 1976), which
also have weak forbidden lines, narrow permitted lines, and strong
\ion{Fe}{2}.  One of the most extreme narrow-line \ion{Fe}{2} QSOs is
PHL~1092 (Bergeron \& Kunth 1980, 1984; Kwan et al. 1995).  Its
\ion{Fe}{2}~$\lambda$4570/H$\beta$ ratio is 5.3, and its line widths
are only $1800$~km~s$^{-1}$ (Bergeron \& Kunth 1984).  Only a weak
X-ray detection of PHL~1092 was made by {\it Einstein} (Wilkes et
al. 1994). Supported by the results of a {\it ROSAT\/}
observation of PHL~1092, we argue that the NLS1s and their QSO analogs
can be understood as a single phenomenon, the ``I~Zw~1'' objects.  In
addition to sharing their extreme X-ray behavior, PHL~1092 might
illuminate fundamental puzzles about this distinctive class.

\section{Observations and Results}

An archival {\it ROSAT\/} PSPC observation of PHL~1092 was reduced in
the standard way using the PROS version 2.3 X-ray analysis package and
XSPEC for spectral fitting.  A summary of the observation is presented
in Table 1.  Only counts in Pulse Invariant (PI) channels 0.11--2.01
keV were used.  A power law fitted the spectrum very well, with photon
index $\Gamma = 4.17^{+ 0.63}_{-0.50}$ (90\% confidence), luminosity
$L_{\rm 0.1 - 2.0 \hskip 0.1cm keV} =$ (5.4 $^{+11.5}_{-3.2}$) $\times
10^{46}$ erg s$^{-1}$ and $N_{\rm H}$ consistent with the Galactic 21
cm value (Murphy et al. 1996).  Although this spectrum is very steep,
it is not unusual for the NLS1 class (Boller et al. 1996; Forster
1996).  Figure~1 shows the fitted PSPC spectrum of PHL~1092, the
residuals from the power-law fit, and the $\chi^{2}$ confidence
contours of the fitted parameters.  Other simple models, such as
blackbody or thermal bremsstrahlung, significantly underpredicted the
Galactic $N_{\rm H}$.  Blackbody plus power-law and bremsstrahlung
plus power-law models did not improve on the single power-law fit, and
more complicated models were not justified by the results of an $F$
test. The rapid X-ray variability of PHL~1092 during the $ROSAT$
observation rules out the possibility of a thermal bremsstrahlung
component providing the bulk of the X-ray luminosity (see \S3.2).

\begin{figure}[tp]
%\figurenum{1}
\epsscale{0.9}
%\epsffile{f1.eps}
\plotone{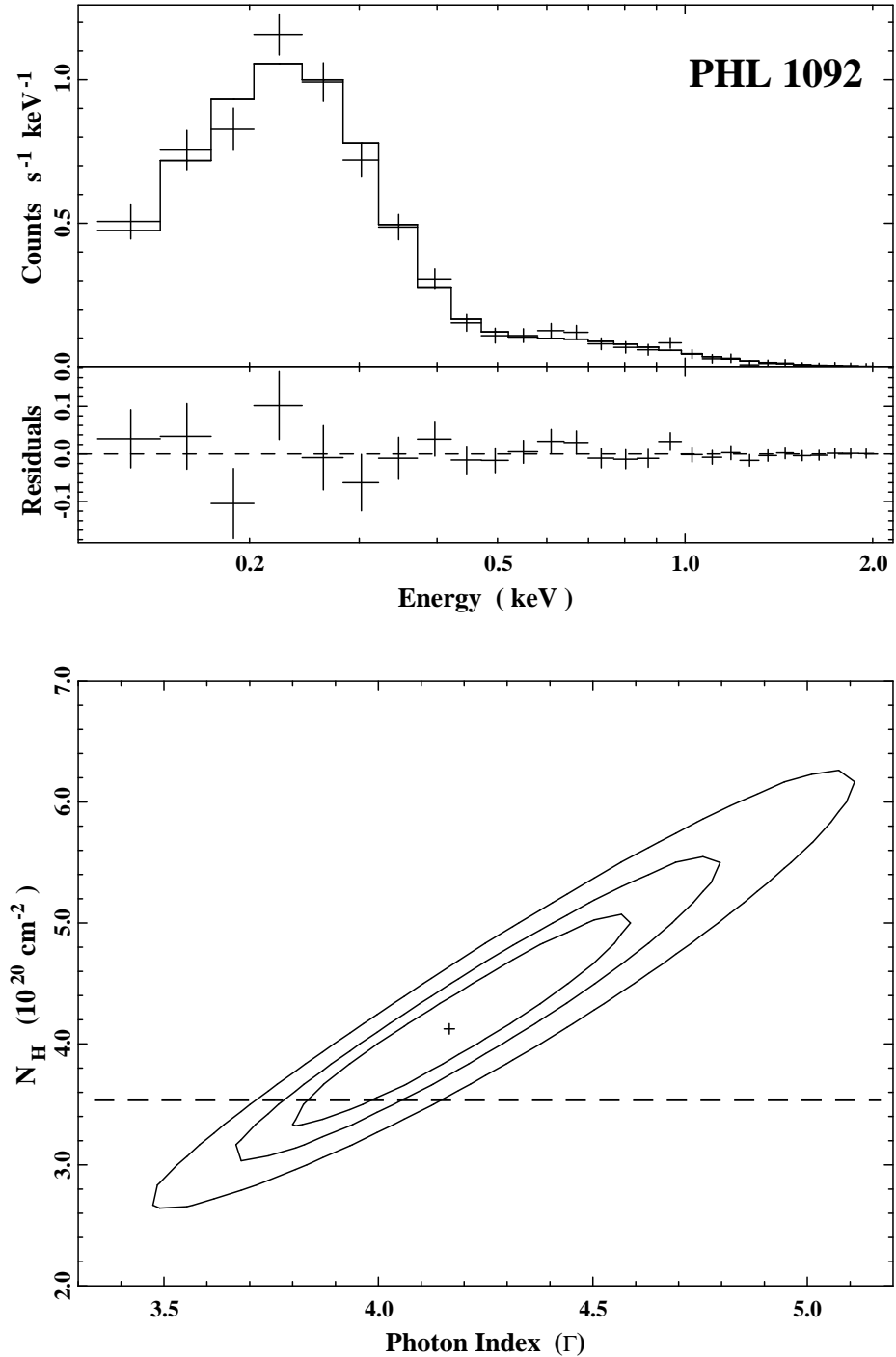}
%\plottwo{fig1ax.eps}{fig1bx.eps}
%\plotfiddle{fig1a.eps}{10in}{270}{80}{80}{0}{0}
\caption{{\it ROSAT \rm} PSPC spectrum of PHL 1092 with best
fitting power-law model.  Also plotted are the residuals from the fit
and $\chi^{2}$ contours for the 68\%, 90\% and 95\% confidence limits
for the fitted paramaters.  The horizontal dashed line represents the
Galactic $N_{\rm H}$ on this line of sight from the 21 cm measurement
of Murphy et al. (1996).}
\end{figure}

\begin{table}[t]
\begin{center}
\begin{tabular}{lcclc}
\noalign{TABLE 1}
\noalign{\vskip 0.2cm}
\noalign{$ROSAT$ PSPC OBSERVATION OF PHL 1092}
\noalign{\vskip 0.4cm}
\tableline
\tableline
\noalign{\vskip 0.2cm}
\noalign{\hskip -2cm Observational details \hskip 4cm 
Power-law model$^{\rm a}$}
\noalign{\vskip 0.2cm}
\tableline
{$\alpha_{(2000)}$ \dotfill} & 
$01^{\rm h} \hskip 0.1cm 39^{\rm m} \hskip 0.1cm 55^{\rm s}.8$ & 
{\hskip 0.5cm} & 
{$\Gamma$ \dotfill} & 4.17 $^{+0.63}_{-0.50}$ \\
{$\delta_{(2000)}$ \dotfill} & 
$06^{\circ} \hskip 0.1cm 19^{\prime} \hskip 0.1cm 21^{\prime\prime}.3$ && 
$A$ (ph cm$^{-2}$ s$^{-1}$ keV$^{-1}$) & (2.52 $^{+0.62}_{-0.61}$) 
$\times 10^{-4}$\\
{$z$ \dotfill} & 0.396 && 
{$N_{\rm H}$ (cm$^{-2}$) \dotfill} & (4.13 $^{+1.41}_{-1.10}$) 
$\times 10^{20}$ \\ 
{$N_{\rm H I}^{^{\rm Gal}}$ (cm$^{-2})\ ^{\rm b}$ \dotfill}& 
($3.53 \pm 0.10$) $\times 10^{20}$ && 
{$\chi^{2}_{\nu}$ ($\nu$)  \dotfill}& 0.541 (25) \\
{Date \dotfill} & 1992 Jan 19--22 && 
$F_{\rm X}$ (ergs cm$^{-2}$ s$^{-1})\ ^{\rm c}$ & $1.93 \times 10^{-12}$ \\
{T$_{\rm obs}$ (s) \dotfill} & 7613 && 
{$L_{\rm X}$ (ergs s$^{-1})\ ^{\rm d}$ \dotfill} & (5.42 $^{+11.50}_{-3.16}$) 
$\times 10^{46}$ \\
Counts s$^{-1\ {\rm c}}$ & 0.283 $\pm$ 0.008 \\
\noalign{\vskip 0.2cm}
\tableline
\end{tabular}

\end{center}

\tablenotetext{}{$^{\rm a}$ Errors are 90\% confidence.}

\tablenotetext{}{$^{\rm b}$ Murphy et al. (1996).}

\tablenotetext{}{$^{\rm c}$ 0.1 -- 2.0 keV observed frame.}

\tablenotetext{}{$^{\rm d}$ 0.1 -- 2.0 keV rest frame, 
assuming $H_0 = 50$ km s$^{-1}$ Mpc$^{-1}, q_0=0$.}

\end{table}

\begin{figure}[t]
%\figurenum{2}
%\epsscale{0.5}
%\epsffile{f1.eps}
%\plotone{sgi9289.eps}
\plotone{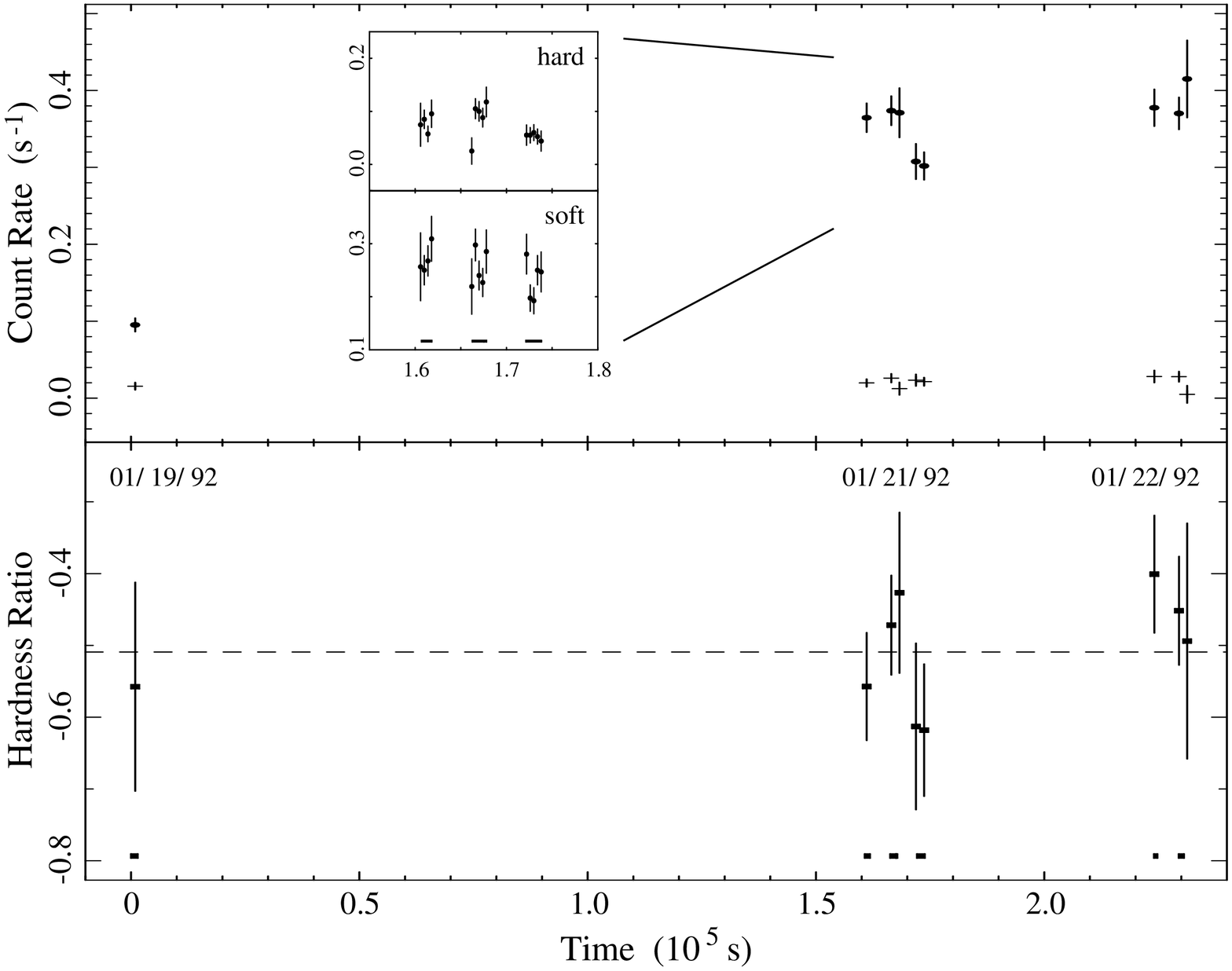}
%\plotfiddle{fig2t.eps}{3in}{0}{70}{70}{-200}{-200}
\caption{Top: {\it ROSAT\/} PSPC light curve during the
1992 January observation of PHL~1092. Background-subtracted source
counts in 1800 second bins (dark ovals) and associated 1$\sigma$
errors are shown.  Crosses indicate the background count rate in the
same size region as the source.  The inset shows the behavior of the
hard (H = 0.4--2.0 keV) and soft (S = 0.1--0.4 keV) events on January
21.  Bottom: The hardness ratio (H -- S)/(H + S) and 1$\sigma$
errors. The rectangles along the bottom represent the length of each
good time interval during the observation.  The horizontal dashed line
marks the average hardness ratio ($-0.51 \pm 0.08$).}
\end{figure}

Figure 2 shows the light curve of this PSPC observation, which is
spread over three days.  There was a highly significant increase in
flux by a factor of 4 sometime during the intial two-day period,
followed by a decrease of about 20\% in 6000~s, albeit at a lower
level of significance.  Small changes in the hardness ratio possibly
occurred as well.  Because the luminosity is so large, $\approx 5
\times 10^{46}$~ergs~s$^{-1}$, these X-ray variations stretch the
limits that can be achieved by an isotropic source powered by
accretion onto a non-rotating black hole as expressed by the relation
$\eta\,>\,5 \times 10^{-43}\,\Delta L/\Delta t$ (Fabian 1979; Fabian
\& Rees 1979), where $\Delta t$ is the time scale and $\Delta L$ the
amplitude of variability.  The required radiative efficiency $\eta$
would be $>0.13$ for the initial (unresolved) increase by a factor of
4, and possibly as high as 0.61 if the subsequent variation in 6000~s
is real.  Values of $\eta$ larger than 0.057, the limit for an
accretion disk in the Schwarzschild metric, require either accretion
onto a Kerr black hole, or modification of the assumptions implicit in
the Fabian-Rees relation, for example invoking anisotropic emission.
The addition of a blackbody of temperature $kT=93$~eV to the power-law
model, although not justified statistically by an $F$ test, reduces
the X-ray luminosity to $\sim 2.1~\times 10^{46}$~ergs~s$^{-1}$, but
this still requires the radiative efficiency to exceed 0.062 for the
initial increase. The blackbody plus power-law model is very poorly
constrained.  The 90\% confidence limit on the parameters of the
single power-law model indicate a similar lower limit to the X-ray
luminosity, $L_{\rm 0.1 - 2.0 \hskip 0.1cm keV} >
2.2~\times~10^{46}$~ergs~s$^{-1}$.

In addition to its rapid variability, PHL~1092 was a much brighter
{\it ROSAT\/} X-ray source than would have been expected from the weak
{\it Einstein} IPC detection in July~1979.  In a 6389 s IPC
observation, only $30 \pm 11$ counts were collected, corresponding to
a probable $0.2-4.5$~keV luminosity of $\sim 3 \times
10^{44}$~ergs~s$^{-1}$ (Wilkes et al. 1994).  Although part of the
factor of $\sim 200$ discrepancy between the {\it Einstein} and {\it
ROSAT\/} peak luminosities is undoubtedly due to the extremely steep
spectral index of PHL~1092 and the different bandpasses of the two
instruments, the {\it Einstein} count rate is still a factor of 21
smaller than would be predicted by folding the {\it ROSAT\/} fitted
spectrum through the IPC response (using PIMMS, see Table 2).

\newpage
\section {Discussion}

\subsection {Long-term Variability}

Long-term X-ray variability by more than a factor of 10 is rare among
AGNs, and the most dramatic examples seem to be members of the I~Zw~1
class.  Table 2 is a compilation of such reports from the recent
literature, involving intercomparisons among {\it HEAO 1}, {\it
Einstein}, {\it EXOSAT\/}, and {\it ROSAT}.  All but E1615+061 satisfy
the I~Zw~1 definition.  We have also added the narrow-line QSO
PG~1404+226 (FWHM H$\beta$ = 880~km~s$^{-1}$, \ion{Fe}{2}/H$\beta$
=1.01, Boroson \& Green 1992) based on the {\it ROSAT\/} data of
Ulrich \& Molendi (1996), which indicates a brightening by a factor of
13 from the {\it Einstein} flux (assuming the PSPC photon index of
3.57 applies to both observations).  One case we have {\it not}
included is that of PG~1211+143, as we could not verify the claim of
Yaqoob et al. (1994) that the {\it ROSAT\/} X-ray flux from 
this narrow-line QSO increased by a factor of $\sim$ 16
since the {\it Einstein} observations.  Instead, we find that for the
steep {\it ROSAT\/} PSPC spectrum ($\Gamma = 3.10 \pm 0.13$, Forster
1996), which is

\begin{table}[ht]
\begin{center}
\begin{tabular}{lcccccc}
\noalign{TABLE 2}
\noalign{\vskip 0.2cm}
\noalign{LONG-TERM X-RAY VARIABILITY IN AGNs}
\noalign{\vskip 0.4cm}
\tableline
\tableline
\noalign{\vskip 0.2cm}
\noalign{$Einstein$ \hskip 3.3cm $ROSAT$ \hskip 3cm}
\quad Name & Date & Counts s$^{-1}$ &  Date & Counts s$^{-1}$ &
$\Delta F/F\ ^{\rm a}$ \\
&& (0.16 -- 3.5 keV) & & (0.1 -- 2.0 keV) \\
\noalign {\vskip 0.1cm}
\tableline
PHL 1092 \dotfill & 1979 July & $4.6 \times 10^{-3}$   & 1992 Jan & 
0.28 & 20.9 \\
PG 1404+226 & 1981 Jan & $1.6 \times 10^{-2}$   & 1991 July &
0.63 & 13.1 \\
\tableline
\\
\noalign{RASS \hskip 4cm $ROSAT$ \hskip 3cm }
\quad Name & Date & Counts s$^{-1}$ & Date & Counts s$^{-1}$ &
$\Delta F/F$  \\
\tableline
WPVS 007$\ ^{\rm b}$ \dotfill & 1990 Nov & 0.98  &
1993 Nov & $2.6 \times 10^{-3}$ & $-377$ \\
IC 3599$\ ^{\rm c}$ \dotfill & 1990 Dec & 4.90 &
1991 Dec & $6.4 \times 10^{-2}$ & $-76$ \\
\tableline
\\
\noalign{{\it HEAO 1} A-2 \hskip 3.2cm $EXOSAT$ \hskip 3cm}
\quad Name & Date & $L_{\rm X}$ (ergs s$^{-1}$) & 
Date & $L_{\rm X}$ (ergs s$^{-1}$) & $\Delta L/L$ \\
\tableline
E1615+061$\ ^{\rm d}$ \dotfill & 1977 Aug & $2.4 \times 10^{45}$  & 
1985 July & $1.3 \times 10^{43}$ & $-185$ \\
\tableline
\end{tabular}
 
\end{center}

\tablenotetext{}{$^{\rm a}$ Change in flux referenced to the same bandpass
 using PIMMS.}
\tablenotetext{}{$^{\rm b}$ Grupe et al. (1995b).}
\tablenotetext{}{$^{\rm c}$ Grupe et al. (1995a).}
\tablenotetext{}{$^{\rm d}$ 0.1--6.0 keV, Piro et al. (1988).}

\end{table}

\noindent similar to the {\it Einstein} IPC result ($\Gamma =
3.0 \pm 0.7$ in PHA channels 1--6 only, Bechtold et al. 1987), the
change in intrinsic luminosity in this object is only a factor of 3.

The mechanism of the extreme X-ray variability remains unknown.
Although part of the reason for such dramatic variability in
steep-spectrum objects could be changes in spectral shape that move
the bulk of the bolometric luminosity in and out of the observed
bandpasses (e.g. the tail of an accretion disk spectrum, Marshall et
al. 1996), there is no strong evidence that this is a sufficient or
general explanation.  The spectrum of IC~3599 was steeper in its low
state (Brandt et al. 1995; Grupe et al. 1995a), while the opposite was
the case for E1615+061 (Piro et al. 1988).  The object with the
largest amplitude of variability, WPVS~007, showed no significant
change in hardness ratio (Grupe et al. 1995b).  A decrease in the
H$\alpha$ luminosity of IC~3599 by a factor of 10 in less than a year
(Mannheim et al. 1996) is perhaps independent evidence that the total
ionizing luminosity of this object does vary by a large factor.  On
the other hand, there were no significant changes in the optical
continuum flux or \ion{Fe}{2} line intensities in PHL~1092 between
optical spectra coincidentally taken in 1979 August (Bergeron \& Kunth
1980, 1984) one month after the {\it Einstein} observation, and in
1992 September (Kwan et al. 1995) eight months after {\it ROSAT\/}
observed the QSO in its X-ray high state.

\subsection {Rapid Variability}

The study of rapid X-ray variability in AGNs has a long tradition
(e.g. Barr \& Mushotzky 1986; Feigelson et al. 1986; Green, McHardy,
\& Lehto 1993; Mushotzky, Done, \& Pounds 1993; Papadakis \& Lawrence
1995).  The most extreme cases of variability outside of blazars are
again seen in I~Zw~1 objects.  Examples are given in Table 3, in which
the first four entries are I~Zw~1 objects, and the last two are
blazars.  IRAS 13224--3809 varied by a factor of 2 on time scales of
800~s (Boller et al. 1993), and the QSO PKS 0558--504 (M$_{V} =
-25.1$, Remillard et al. 1986) showed a hard X-ray flare with a rise
time of 180 seconds during a {\it Ginga} observation in 1989
(Remillard et al. 1991).  These two objects have strong, narrow
\ion{Fe}{2} emission lines (\ion{Fe}{2}/H$\beta$ = 1.83 and 1.56, FWHM
H$\beta$ = 650 and 1500~km~s$^{-1}$, respectively, for
IRAS~13224--3809 and PKS~0558--504), and the {\it ROSAT\/} spectrum
of IRAS 13224--3809 is also very steep ($\Gamma > 4$, Boller et al.
1993).

\begin{table}[t]
\begin{center}
\begin{tabular}{lcrcccl}
\noalign{TABLE 3}
\noalign{\vskip 0.2cm}
\noalign{RAPID CHANGES IN X-RAY LUMINOSITY}
\noalign{\vskip 0.4cm}
\tableline
\tableline
\noalign{\vskip 0.2cm}
\quad Name & Instrument & {$\Delta$L \hskip 0.4cm} & $\Delta$t & 
$\eta$ & Reference \\
& & (ergs s$^{-1}$)  & (s) \\
\noalign {\vskip 0.2cm}
\tableline
PHL 1092 \dotfill & $ROSAT$ & $-7.3 \times 10^{45}$ & 6000 & 0.61 & 1 \\
& '' & $4.1 \times 10^{46}$ & $1.6 \times 10^{5}$ & 0.13 & 1 \\
IRAS 13224--3809 & $ROSAT$ &  $1.5 \times 10^{44}$ & 800 & 0.09 & 2 \\
PKS 0558--504 \dotfill & $Ginga$ & $5.8 \times 10^{44}$ & 180 & 1.6 & 3 \\
3C 279 \dotfill & $Ginga$ & $7.8 \times 10^{45}$ & 2700 & 1.45 & 4 \\
H 0323+022 \dotfill & $Einstein$ & $1.5 \times 10^{45}$ & 30 & 25 & 5 \\
\tableline
\end{tabular}
\end{center}

\tablenotetext{}{References.---(1) This paper;
(2) Boller et al. (1993); (3) Remillard et al. (1991);}

\tablenotetext{}{(4) Makino et al. (1989); (5) Feigelson et al. (1986).}

\end{table}

Estimated lower limits on the efficiency implied by the observed
variations in X-ray flux are also given in Table~3.  The observed
X-ray variability in PHL~1092 is not the most rapid seen among AGNs
but it is unusual in an object of such high X-ray luminosity.  This
has implications for the assumptions usually made about the accretion
process and the generation of X-rays from the inner regions of an
accretion disk.  The Fabian-Rees relation assumes that the flux
released from conversion of mass to energy is isotropically emitted
from a spherically symmetric, homogeneous region where no bulk
relativistic motions occur, while the upper limit of 0.057 for the
efficiency applies to the particular case of a thin accretion disk
around a non-rotating black hole.  In PHL~1092 and similar objects, it
is likely that either the radiation is emitted anisotropically, or the
accretion disk extends to smaller radius as a result of rotation of
the black hole.

Blazars also show rapid variability and high X-ray luminosity, and
many models of their broad-band energy distribution invoke bulk
relativistic motion of the emitting plasma.  It is much less
compelling to apply such a model to I~Zw~1 objects, as their
similarity to blazars doesn't extend much past their X-ray
variability.  I~Zw~1s are generally radio quiet, are weakly polarized
in the optical (Goodrich 1989), and have strong optical emission
lines.  The rapid response of the Balmer lines to the ionizing
continuum in IC~3599 (Mannheim et al. 1996) indicates that dense
emission-line clouds can indeed exist close to the nucleus in these
objects, and argues against the hypothesis that a steep X-ray spectrum
destroys the emission-line clouds through the lack of a two-phase
photoionization equilibrium (Guilbert, Fabian, \& McCray 1983).

\subsection{Narrow-Line Seyferts and QSOs as a Class}

While analyzing the {\it ROSAT} spectra of a large sample of optically
selected I~Zw~1 objects, we discovered a significant ($>$ 99\%
probability) correlation between their intrinsic soft X-ray luminosity
and their X-ray spectral slope.  This correlation is presented in
Figure~3; the data are discussed in detail by Forster (1996).  The
correlation exists across four orders of magnitude in X-ray luminosity
and is highly significant even if PHL~1092 is excluded from the
sample.  The position of PHL~1092 in this diagram argues for it to be
regarded as a high-luminosity but otherwise typical member of the
I~Zw~1 class.  The rather shallow dependence of the photon index over
many orders of magnitude in luminosity argues for a weak dependence of
spectral shape on some fundamental parameter, possibly the mass of the
black hole.

Another interesting question involves the relationship between the
objects discussed here and the elusive ``type 2 QSOs''
(high-luminosity counterparts of Seyfert~2 galaxies).  
Independent of
any orientation-dependent unification schemes, their small
[\ion{O}{3}]/H$\beta$ ratios,

\begin{figure}[bh]
%\figurenum{3}
%\epsscale{0.5}
%\epsffile{f1.eps}
\plotone{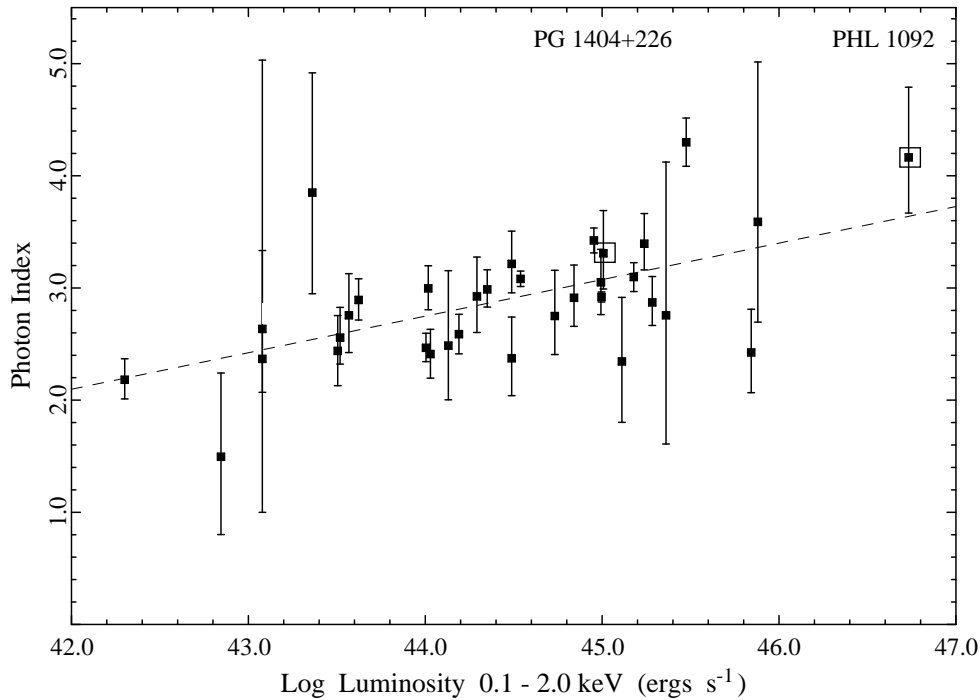}
%\plottwo{fig1ax.eps}{fig1bx.eps}
%\plotfiddle{fig1a.eps}{10in}{270}{80}{80}{0}{0}
\caption{Relationship between intrinsic soft X-ray
luminosity and {\it ROSAT\/} PSPC photon index for optically selected
NLS1s, taken from Forster (1996).  The error bars are the 90\%
confidence limits of the power-law model.  The dashed line is the
linear least-squares regression ${\Delta \Gamma / \Delta {\rm log}\,
L_{\rm X}} = 0.32$.  PHL~1092 and PG~1404+226 are marked with open
boxes.}
\end{figure}
\newpage

\noindent  rapid X-ray variability, and lack of
X-ray obscuration disqualify the I~Zw~1 objects as luminous Seyfert
2s.  Neither does it seem likely that they are the ``pole-on'' parent
population of Seyfert~2 galaxies in unification models, because the
majority of hidden Seyfert~1 galaxies revealed in polarized light do
not have strong permitted \ion{Fe}{2} lines (Tran 1995), while
\ion{Fe}{2} emission is a hallmark of the I~Zw~1 class.

It remains, then, to ask if there are {\it any} high-luminosity
counterparts of Seyfert~2 galaxies.  Limiting the discussion to X-ray
sources, we are aware of only four such candidates.  The oft-mentioned
{\it Einstein} source 1E~0449--184 ($z = 0.338$, Stocke et al. 1982)
is not a secure example because its classification is based solely on
the absence of broad H$\beta$.  Its H$\alpha$ line has never been
examined, and its \ion{Mg}{2} line does appear to be broad (Stocke et
al. 1983).  A more recent candidate, IRAS~20181--2244 ($z = 0.185$,
Elizalde \& Steiner 1994), selected from the {\it ROSAT\/} All-Sky
Survey, was subsequently shown with better spectroscopy to be a I~Zw~1
object (Moran et al. 1996), and not even the most luminous one among
20 similar objects in that survey.  The latest candidates are the {\it
ROSAT\/} source RX J13434+0001 at $z = 2.35$ (Almaini et al. 1995),
and an $ASCA$ source at $z = 0.9$ (AX~J08494+4454, Ohta et al. 1996), but their
faintness and high redshift do not yet allow evaluation in comparable
detail to that of the better studied low-redshift objects.  A fifth
candidate, the hyperluminous {\it IRAS} galaxy P09104+4109
($z~=~0.442$), was at first thought to be the most luminous hidden AGN
in X-rays based on its spectrum from $ASCA$ (Fabian et
al. 1994). However, a subsequent $ROSAT$ HRI image showed that the
bulk of the X-ray emission is extended, and probably all originates in
a massive cooling flow (Fabian \& Crawford 1995). So while there are
many ultraluminous {\it IRAS} galaxies with optical spectra like
Seyfert 2 galaxies, there is little, if any, evidence that they are
luminous X-ray sources.  In any case, it appears that QSOs of the
I~Zw~1 class must be more numerous than any ``type 2 QSOs.''

Finally, one of the oldest unsolved mysteries in the study of AGNs is
the ``\ion{Fe}{2} problem''.  Photoionization models cannot account
for the \ion{Fe}{2}/H$\beta$ ratio in those objects in which it is
largest.  Consequently, mechanical sources of heating have been
postulated.  Kwan et al. (1995) recently proposed a scenario for
PHL~1092 in which \ion{Fe}{2} (and \ion{Fe}{1}) emission is produced
in an accretion disk at low temperature by the dissipation of orbital
energy as the material is forced to corotate with magnetic field lines
tied to a Kerr black hole.  Quite independently, we find evidence from
the X-ray variability of PHL~1092 and similar objects that a Kerr
black hole may be needed to account for their high efficiency.  For
this proposed link to be consistent, however, the \ion{Fe}{2} emitting
region in the outer disk must be largely shielded from view of the
luminous photoionizing source at the nucleus.

\section{Summary}

We analyzed the {\it ROSAT\/} PSPC observation of the luminous
\ion{Fe}{2} QSO PHL~1092, and conclude that it deserves to be
classified as a luminous narrow-line Seyfert~1 galaxy for the
following reasons:

\noindent 1) Its optical emission-line spectrum conforms with the NLS1
classification (Osterbrook \& Pogge 1985; Goodrich 1989).

\noindent 2) Its soft X-ray spectrum is extremely steep, with photon
index $\Gamma = 4.17 ^{+0.63}_{-0.50}$, very similar to the spectrum
of the NLS1 IRAS 13224--3809.

\noindent 3) The large amplitude variability of its X-ray emission
over long time scales is similar to that of WPVS 007 (Grupe et
al. 1995b) and PG 1404+226, both of which are NLS1s with ultra-soft
X-ray spectra and strong \ion{Fe}{2} emission.

\noindent 4) Its rapid X-ray variability requires high efficiency,
similar to the behavior of IRAS~13224--3809, and especially
PKS~0558--504, which are also I~Zw~1 objects.

\noindent We also found a correlation between X-ray spectral index and
X-ray luminosity which may eventually help to constrain models of
accretion in I~Zw~1 objects.  The possibility that rotating black
holes may be implicated in the explanation of both the unusual X-ray
and optical properties of this class is suggested.

\acknowledgements

This work was support by NASA grant NAG 5-1935 and has made use of the
NASA/IPAC Extragalactic Database (NED) which is operated by the Jet
Propulsion Laboratory, California Institute of Technology, under
contract with the National Aeronautics and Space Administration.  This
research has also made use of data obtained through the High Energy
Astrophysics Science Archive Research Center Online Service, provided
by the NASA-Goddard Space Flight Center.  This paper is contribution
No. 587 of the Columbia Astrophysics Laboratory.

%\newpage

%\figcaption[fig1.ps]{{\it ROSAT\/} PSPC spectrum of PHL 1092 with best
%fitting power-law model.  Also plotted are the residuals from the fit
%and $\chi^{2}$ contours for the 68\%, 90\% and 95\% confidence limits
%for the fitted paramaters.  The horizontal dashed line represents the
%Galactic $N_{\rm H}$ on this line of sight from the 21 cm measurement
%of Murphy et al. (1996).\label{fig1}}

%\figcaption[fig2.ps]{Top: {\it ROSAT\/} PSPC light curve during the
%1992 January observation of PHL~1092. Background-subtracted source
%counts in 1800 second bins (dark ovals) and associated 1$\sigma$
%errors are shown.  Crosses indicate the background count rate in the
%same size region as the source.  The inset shows the behavior of the
%hard (H = 0.4--2.0 keV) and soft (S = 0.1--0.4 keV) events on January
%21.  Bottom: The hardness ratio (H -- S)/(H + S) and 1$\sigma$
%errors. The rectangles along the bottom represent the length of each
%good time interval during the observation.  The horizontal dashed line
%marks the average hardness ratio ($-0.51 \pm 0.08$).\label{fig2}}

%\figcaption[fig3.ps]{Relationship between intrinsic soft X-ray
%luminosity and {\it ROSAT\/} PSPC photon index for optically selected
%NLS1s, taken from Forster (1996).  The error bars are the 90\%
%confidence limits of the power-law model.  The dashed line is the
%linear least-squares regression ${\Delta \Gamma / \Delta {\rm log}\,
%L_{\rm X}} = 0.32$.  PHL~1092 and PG~1404+226 are marked with open
%boxes.\label{fig3}}

\end{document}